\documentclass[11pt]{article}
\usepackage{cite}
\usepackage{graphicx}
\usepackage{amsfonts,amsmath,amssymb}
\usepackage{color}
\usepackage{braket}
\usepackage[]{epsfig}
\usepackage{enumitem}
\usepackage{epstopdf}
\usepackage{multirow}
\usepackage{overpic}
\usepackage{colortbl}
\usepackage{tabularx}
\usepackage{url}
\usepackage{color}
\usepackage{bm}
\usepackage{units}
\usepackage[colorlinks,
            linkcolor=blue,       
            anchorcolor=blue,  
            citecolor=blue]{hyperref}

\newcommand{\ee}{e^+e^-}

\newcommand{\dzero}{D^{0}}

\newcommand{\dzerobar}{\bar{D}^{0}}

\newcommand{\B}{\mathcal{B}}

\begin{document}

\begin{center} \begin{LARGE} \textbf{
Charm Physics at the Super $\tau$-Charm Factory} \end{LARGE} \end{center}

\begin{center}
{\large Hai-Yang Cheng,}$^a$~\footnote{E-mail: phcheng@phys.sinica.edu.tw}
{\large Xiao-Rui Lyu,}$^b$~\footnote{E-mail: xiaorui@ucas.ac.cn} 
{\large Zhi-Zhong Xing}$^{b,c}$~\footnote{E-mail: xingzz@ihep.ac.cn} \\ \vspace{0.3cm}
$^a$ Institute of Physics, Academia Sinica, Taiwan 11529, China \\
$^b$ University of Chinese Academy of Sciences, Beijing 100049, China
 \\
$^c$  Institute of High-Energy Physics, Beijing 100049,  China \\
 \vspace{0.5cm}
\end{center}

\abstract{
The high-luminosity Super $\tau$-Charm Factory (STCF) will be a crucial facility for charm-physics research, particularly for the precise measurement of electroweak parameters, measuring $D^0$-$\bar{D}^0$ mixing parameters, investigating Charge-Parity (CP) violation within the charm sector, searching for the rare and forbidden decays of charmed hadrons, and addressing other foundational questions related to charmed hadrons. With the world’s largest charm-threshold data, the STCF aims to achieve high sensitivity in studying the strong phase of neutral $D$ mesons using quantum correlation, complementing studies at LHCb and Belle II, and contributing to the understanding of CP violations globally. The STCF will also enable world-leading precision in measuring the leptonic decays of charmed mesons and baryons, providing constraints on the Cabibbo-Kobayashi-Maskawa matrix and strong-force dynamics. Additionally, the STCF will explore charmed hadron spectroscopy. The advanced detector and clean experimental environment of the STCF will enable unprecedented precision, help address key challenges in the Standard Model, and facilitate the search for potential new physics.
}

\section{Introduction}

The identification of the charm quark in 1974 marked a significant milestone in the evolution of particle physics and formulation of the Standard Model (SM). The high-luminosity Super $\tau$-Charm Factory (STCF), as envisioned in \cite{Peng:2020orp,Wang:2021pus,Lyu:2021tlb}, will provide a pivotal platform for advancing our understanding of the SM and exploring new physics. The STCF is designed to produce approximately $10^9$--$10^{10}$ quantum-coherent $D^0 \bar{D}^0$ meson pairs and $D_{(s)}^+$ mesons, along with over $10^8$ $\Lambda_c^+$ baryons and heavier charmed baryons, all within a low-background environment. Building on the success of the BESIII experiment \cite{Li:2021iwf}, the STCF will offer unparalleled opportunities.

Notably, the STCF will serve as a distinctive instrument for determining the Cabibbo-Kobayashi-Maskawa (CKM) matrix elements $V_{cd}$ and $V_{cs}$, measuring $D^0$-$\bar{D}^0$ mixing parameters, investigating CP violations within the charm sector, searching for rare and forbidden decays of charmed hadrons, and addressing other foundational questions related to charmed hadrons. The precision of many key measurements at the STCF is expected to be limited by systematic uncertainties, thereby necessitating a cutting-edge detector system that excels in particle identification, particularly in distinguishing between various charged particles, detecting low-momentum charged particles, and measuring photons, as described in Ref.~\cite{Shi:2020nrf}.

\section{Charmed meson}

\subsection{$D^+_{(s)}$ leptonic decays}

\begin{table*}[th]
\centering
\caption{\label{tab:pure_LP}\small
Precision of studies on $D^+_{(s)} \to \ell^+ \nu_\ell $ at BESIII, projected for STCF, and estimated for Belle II. The precisions with marker $^*$ are scaled from the BESIII measurement in Ref.~\cite{BESIII:2024vlt} with 7.9 fb$^{-1}$ at 3.773 GeV. For Belle II, we assume that the systematic uncertainties are halved compared to those of Belle.}
\begin{tabular}{lcccc} \hline\hline
\multicolumn{1}{c}{} & BESIII & STCF & Belle II \\ \hline
Luminosity & 20 fb$^{-1}$ at 3.773 GeV & 1 ab$^{-1}$ at 3.773 GeV & 50 ab$^{-1}$ at $\Upsilon(nS)$ \\ \hline
${\mathcal B}(D^+ \to \mu^+ \nu_\mu)$ & $2.0\%_{\text{stat}}\,1.0\%_{\text{syst}}$~\cite{BESIII:2024kvt}
 & $0.28\%_{\text{stat}}$ & -- \\
$f_{D^+}$ (MeV) & $1.0\%_{\text{stat}}\,0.6\%_{\text{syst}}$~\cite{BESIII:2024kvt}  & $0.15\%_{\text{stat}}$ & -- \\
$|V_{cd}|$ & $1.0\%_{\text{stat}}\,0.6\%_{\text{syst}}$~\cite{BESIII:2024kvt} & $0.15\%_{\text{stat}}$ & -- \\
${\mathcal B}(D^+ \to \tau^+ \nu_\tau)$ & $7.0\%_{\text{stat}}^*\,4.0\%_{\text{syst}}^*$~\cite{BESIII:2024vlt} & $0.41\%_{\text{stat}}$ & -- \\
$\displaystyle \frac{{\mathcal B}(D^+ \to \tau^+ \nu_\tau)}{{\mathcal B}(D^+ \to \mu^+ \nu_\mu)}$ & $7.3\%_{\text{stat}}^*\,4.1\%_{\text{syst}}^*$~\cite{BESIII:2024vlt} & $0.50\%_{\text{stat}}$ & -- \\ \hline \hline
Luminosity & 7.3 fb$^{-1}$ around 4.178 GeV & 1 ab$^{-1}$ at 4.009 GeV & 50 ab$^{-1}$ at $\Upsilon(nS)$ \\ \hline
${\mathcal B}(D^+_s \to \mu^+ \nu_\mu)$ & $2.0\%_{\text{stat}}\,1.6\%_{\text{syst}}$~\cite{BESIII:2023cym} & $0.30\%_{\text{stat}}$ & $0.8\%_{\text{stat}}\,1.8\%_{\text{syst}}$ \\
$f_{D^+_s}$ (MeV) & $1.0\%_{\text{stat}}\,0.9\%_{\text{syst}}$~\cite{BESIII:2023cym}
 & $0.15\%_{\text{stat}}$ & -- \\
$|V_{cs}|$ & $1.0\%_{\text{stat}}\,0.9\%_{\text{syst}}$~\cite{BESIII:2023cym}
 & $0.15\%_{\text{stat}}$ & -- \\ \hline
${\mathcal B}(D^+_s \to \tau^+ \nu_\tau)$ & $1.3\%_{\text{stat}}\,1.5\%_{\text{syst}}$~\cite{BESIII:2023ukh,BESIII:2023fhe} & $0.24\%_{\text{stat}}$ & $0.6\%_{\text{stat}}\,2.7\%_{\text{syst}}$ \\ 
$f_{D^+_s}$ (MeV) & $0.7\%_{\text{stat}}\,0.8\%_{\text{syst}}$~\cite{BESIII:2023ukh,BESIII:2023fhe}  & $0.11\%_{\text{stat}}$ & -- \\
$|V_{cs}|$ & $0.7\%_{\text{stat}}\,0.8\%_{\text{syst}}$~\cite{BESIII:2023ukh,BESIII:2023fhe}  & $0.11\%_{\text{stat}}$ & -- \\ \hline
$\overline f^{\mu \&\tau}_{D^+_s}$ (MeV) & $0.6\%_{\text{stat}}\,0.7\%_{\text{syst}}$ & $0.09\%_{\text{stat}}$ & $0.3\%_{\text{stat}}\,1.0\%_{\text{syst}}$ \\
$|\overline V_{cs}^{\mu \&\tau}|$ & $0.6\%_{\text{stat}}\,0.7\%_{\text{syst}}$ & $0.09\%_{\text{stat}}$ & -- \\ \hline \hline
$\displaystyle \frac{{\mathcal B}(D^+_s \to \tau^+ \nu_\tau)}{{\mathcal B}(D^+_s \to \mu^+ \nu_\mu)}$ & $2.4\%_{\text{stat}}\,2.2\%_{\text{syst}}$ & $0.38\%_{\text{stat}}$ & $0.9\%_{\text{stat}}\,3.2\%_{\text{syst}}$ \\ \hline \hline
\end{tabular}
\end{table*}

The direct determination of the CKM matrix elements, $ |V_{cd}| $ and$ |V_{cs}| $, is one of the most important objectives in charm physics. These parameters dictate the rates of leptonic $D^+$ and $D^+_s $ decays and are pivotal for testing the unitarity of the CKM matrix. The precise measurement of these quantities is a key focus of the STCF project.

The most accurate method for determining $ |V_{cd}| $ and $ |V_{cs}| $ at the STCF involves pure leptonic decay $ D_{(s)}^+ \to \ell^+ \nu_\ell $ ($\ell = e, \mu, \tau$) instead of semileptonic decay, which suffers from larger uncertainties owing to calculations of form factors in Lattice Quantum Chromodynamics (LQCD). The product of the decay constant $f_{D_{(s)}^+}$ and the CKM element $|V_{cd(s)}|$, can be directly inferred from the measurement of the partial widths of $D_{(s)}^+ \to \ell^+ \nu_\ell$. With $f_{D_{(s)}^+}$ input from LQCD, $|V_{cd(s)}|$ can then be derived. 
Moreover, with $|V_{cd(s)}|$ input from global CKM fits, such as CKMfitter~\cite{Charles:2004jd,CKMfitter_web} and UTfit~\cite{Bona:2005vz,utfit_web}, the decay constant $f_{D_{(s)}}$ can be obtained.
The recent determinations of $|V_{cs(d)}|$ and $ f_{D^+_{(s)}} $~\cite{BESIII:2023fhe,BESIII:2023cym,BESIII:2023ukh,BESIII:2024kvt,BESIII:2024vlt} at BESIII, as well as projected precisions at the STCF~\cite{Liu:2021qio,Li:2021ala} are listed in Table~\ref{tab:pure_LP}. Notably, for $ {\mathcal B}(D^+ \to \tau^+ \nu_\tau) $, several decay channels of $\tau^+$ are combined to enhance statistical sensitivity.

Systematic uncertainties at the STCF are being optimized to a subleading level, with statistical uncertainties expected to be considerably less than 0.5\%. Thus, reducing the relevant systematic uncertainties from the background and fitting becomes crucial. Therefore, studies on $D_s^+ \to \ell^+ \nu_\ell $ at 4.009 GeV are optimal. The current LQCD calculations yield $ f_{D^+} = 212.7 \pm 0.6 $ MeV, $ f_{D_s^+} = 249.9 \pm 0.4 $ MeV, and $ f_{D_s^+}/f_{D^+} = 1.1749 \pm 0.0016 $ with a precision of approximately 0.2\%~\cite{Bazavov:2017lyh}. These precisions are expected to be below 0.1\% when the STCF is ready for use; thus, improving the systematic uncertainties to match this level is imperative. The efficiencies of muon and electron identification will be critical for reducing the total uncertainty to 0.1\%. 

Precise measurements of the semi-leptonic branching fraction (BF) for $ D_{(s)} \to h \ell^+ \nu_\ell $ can be used to calibrate the LQCD form-factor calculations, where $h$ is a charmless hadron. This can be achieved by using the input values of $|V_{cd(s)}|$ derived from global CKM fits~\cite{CKMfitter_web,utfit_web}.
Calibration of the LQCD form factors is essential for improving the accuracy of theoretical predictions and reducing systematic uncertainties in the determination of CKM matrix elements and other fundamental SM parameters.
In the case of $D_{(s)} \to V(h_1 h_2) \ell^+ \nu_\ell$, where $V$ denotes a vector meson decaying into hadrons $h_1$ and $h_2$, the construction of triple-product T-odd observables enables a high-precision test of time-reversal (T) invariance~\cite{Belanger:1991vx}. Such observables are sensitive probes of CP violation mechanisms beyond the SM and can reveal signatures of new physics, such as those involving multi-Higgs doublets or leptoquarks~\cite{Grossman:2003qi}.

Reference~\cite{Wang:2019wee} suggests that the combined measurements of $D \to K_1(1270) \ell^+ \nu_\ell$ and $B \to K_1(1270) \gamma$ can be used to determine the photon polarization in $b \to s \gamma$ transitions. This approach provides a clean method for probing right-handed couplings, which may be indicative of new physics. A feasibility study indicated that with approximately 60,000 signal events of $D^0 \to K_1(1270)^- e^+ \nu$ collected with 1 ab$^{-1}$ of data at 3.773 GeV, a statistical sensitivity of $1.5 \times 10^{-2}$ can be achieved for the up-down asymmetry ratio at STCF~\cite{Fan:2021mwp}. This sensitivity is significant and demonstrates the potential of the STCF to obtain precise measurements that may lead to new insights into the nature of CP violations and the presence of new physics beyond the SM.

Lepton flavor universality (LFU) is a cornerstone principle in the SM, positing that three generations of leptons---electrons, muons, and tau leptons---interact with other particles with the same strength. Testing the LFU is a powerful way to search for new physics beyond the SM, as any deviation from the SM predictions may indicate the presence of new particles or interactions not considered in the current framework~\cite{prd91_094009}.
In the context of charmed meson leptonic decay, LFU can be tested by comparing the rates of charm weak decays involving different leptons. Specifically, the ratio $R_{D^+_{(s)}}$ of the partial widths for $D^+_{(s)} \to \tau^+ \nu_\tau$ and $D^+_{(s)} \to \mu^+ \nu_\mu$ is given by $R_{D^+_{(s)}}= [m^2_{\tau^+} (m^2_{D^+_{(s)}} - m^2_{\tau^+} )^2]/[m^2_{\mu^+} (m^2_{D^+_{(s)}} - m^2_{\mu^+} )^2]$.
Using the world average values of the masses of leptons and $D^+_{(s)}$ mesons from the Particle Data Group (PDG)~\cite{PDG}, $R_{D^+} = 2.67 \pm 0.01$ and $R_{D^+_s} = 9.75 \pm 0.01$ are predicted in the SM.
The measurements of $R_{D^+_{(s)}}$ by the BESIII collaboration reported values of $2.49 \pm 0.31$~\cite{BESIII:2024vlt} and $10.05 \pm 0.35$~\cite{BESIII:2023cym} for $D^+$ and $D^+_s$, respectively, which are consistent with the SM predictions. However, these measurements are statistically limited.
At the STCF, the statistical precision of $R_{D_{(s)}^{+}}$ is expected to match the uncertainties of the SM predictions, as indicated in Table~\ref{tab:pure_LP}. This enhanced precision will enable a more stringent test of the LFU~\cite{Liu:2021qio,Li:2021ala}, potentially revealing deviations that may indicate new physics, such as the presence of a charged Higgs boson in a two-Higgs-doublet model that interferes with the SM amplitude involving a $W^\pm$ boson~\cite{prd91_094009}.

In addition, the semileptonic decay of charmed mesons, involving muons and electrons, offers another avenue for testing LFU. In particular, the ratios of the partial widths of $D_{(s)} \to h \mu^+ \nu_\mu$ to those of $D_{(s)} \to h e^+ \nu_e$ at different $q^2$ intervals provide a test complementary to those utilizing tauonic decay.
Based on 2.93 fb$^{-1}$ of data at 3.773 GeV, the BESIII collaboration reported precise measurements of the ratios
${\mathcal B}(D^0\to\pi^-\mu^+\nu_\mu)/{\mathcal B}(D^0\to\pi^-e^+\nu_e)=0.922\pm0.030\pm0.022$
and ${\mathcal B}(D^+\to\pi^0\mu^+\nu_\mu)/{\mathcal B}(D^+\to\pi^0e^+\nu_e)=0.964\pm0.037\pm0.026$~\cite{bes3_pimuv}.
These results are consistent with SM predictions, within $1.7\sigma$ and $0.5\sigma$, respectively~\cite{bes3_pimuv}. With the available 20~fb$^{-1}$ of data at 3.773~GeV at BESIII, these comparisons are constrained by the improved LQCD calculation~\cite{FermilabLattice:2022gku}, in which systematic uncertainties become the dominant issue. With the anticipated 1 ab$^{-1}$ of data collected at 3.773 GeV at the STCF, the statistical uncertainty will be reduced to an unprecedented level of 0.2\%, and the control of systematic uncertainties will be crucial. In addition, with the detailed statistics of the STCF data, the test on LFU can be implemented in a fine and thorough manner, such as at different $q^2$ intervals~\cite{bes3_kmuv}. 

\subsection{$\dzero$-$\dzerobar$ mixing and CP violation}

The phenomenon of meson-antimeson mixing, particularly $D^0$-$\bar{D}^0$ mixing, represents a fascinating area of study in the field of particle physics. Unlike the $B$-meson and Kaon systems, CP violations in $D$-meson mixing have not yet been observed. STCF offers a prime opportunity to study $D^0$-$\bar{D}^0$ mixing and CP violations owing to its high luminosity and clean experimental conditions.

The mass eigenstates of the $D^0$ and $\bar{D}^0$ mesons are expressed as
$|D_1\rangle = p |D^0\rangle + q |\bar{D}^0\rangle$ and $|D_2\rangle = p |D^0\rangle - q |\bar{D}^0\rangle $,
where $|p|^2 + |q|^2 = 1$. The $D^0$-$\bar{D}^0$ mixing parameters are defined as $x \equiv (M^{}_2 - M^{}_1)/\Gamma$ and $y \equiv (\Gamma^{}_2 - \Gamma^{}_1)/(2\Gamma)$, where $M^{}_{1,2}$ and $\Gamma^{}_{1,2}$ are the masses and widths of $D^{}_{1,2}$, respectively. The average width $\Gamma$ and mass $M$ are given by $\Gamma \equiv (\Gamma^{}_1 + \Gamma^{}_2)/2$ and $M \equiv (M^{}_1 + M^{}_2)/2$, respectively.
The $D^0$-$\bar{D}^0$ system is unique because it mixes via intermediate states with down-type quarks, and the mixing parameters $x$ and $y$ are notoriously difficult to calculate in the SM owing to large long-distance uncertainties. Theoretical estimates suggest $x\sim y\sim \sin^2\theta^{}_{\rm C} \times [{\rm SU(3) ~ breaking}]^2$, which yields $x\lesssim y$ and $10^{-3} < |x| < 10^{-2}$~\cite{Falk2004}.
Global fits to experimental data by the Heavy Flavor Averaging Group yield $3.2 \times 10^{-3} < x < 4.9 \times 10^{-3}$ and $6.0 \times 10^{-3} < y < 6.9 \times 10^{-3}$ at the 95\% confidence level~\cite{Amhis:2019ckw,hflav_web}, consistent with theoretical predictions.
At the STCF, more precise measurements of $x$ and $y$ can be achieved. Although these measurements may not clarify long-distance effects, they can help detect small CP-violating effects in neutral $D$-meson decays and mixing~\cite{SuperB}. 

The charm sector is a precision laboratory for exploring possible new physics because SM-induced CP-violating asymmetries in $D$-meson decays typically range from $10^{-4}$ to $10^{-3}$~\cite{Xing:2007zz}.
The CP-violating asymmetries in singly Cabibbo-suppressed $D$-meson decays are expected to be larger than those in the Cabibbo-favored and doubly Cabibbo-suppressed decays~\cite{SuperB}. Three types of CP-violating effects exist in $D$-meson decays~\cite{Falk2004}: CP violation in $D^0$-$\bar{D}^0$ mixing, direct CP violation in decay, and CP violation from the interplay of decay and mixing.
In addition, CP violations can be induced by
$K^0$-$\bar{K}^0$ mixing in $D$ decays involving $K^{}_{\rm S}$ or
$K^{}_{\rm L}$ with a typical magnitude of $2
{\rm Re}(\epsilon^{}_K) \simeq 3.3 \times 10^{-3}$. This may be 
comparable to or even greater than the {\it charmed} CP-violating
effects~\cite{Xing:1995jg,Yu:2017oky}.
Among the many efforts to search for a charm CP violation, the LHCb collaboration first observed a CP violation in the combined $D^0 \to K^+K^-$ and $D^0 \to \pi^+\pi^-$ decays with a significance of 5.3$\sigma$, giving a time-integrated CP-violating asymmetry of $\Delta a_{CP} = (-0.154\pm0.029)\%$~\cite{Aaij:2019kcg}, which is primarily attributed to a direct CP violation in the charm-quark decay~\cite{Saur:2020rgd}.
This is consistent with some theoretical estimates in the SM~\cite{Cheng,Li,Grinstein,Gronau,Italy,Italy2,Li:2019hho,Grossman:2019xcj}.
However, the latter decay suffers from significant uncertainties.
STCF will achieve a $10^{-4}$ level of sensitivity in searching for CP violations in charm meson decays, with kinematic constraints enhancing its capability to study CP-violating asymmetries in multibody $D$-decays~\cite{Bigi:2011em}. 
The CKM mechanism of CP violation within the SM falls short of explaining the observed matter-antimatter asymmetry in the universe by over 10 orders of magnitude~\cite{Morrissey:2012db}. This significant discrepancy strongly motivates the search for new, undiscovered sources of CP violation, potentially involving both quark and lepton flavors. In this context, the charm-quark sector presents a promising avenue for exploration.

The STCF will offer a unique opportunity to study $D^0$-$\bar{D}^0$ mixing and CP violations by leveraging the quantum coherence of $D^0$ and $\bar{D}^0$ mesons produced near the threshold. At the $D^0\bar{D}^0$  threshold at 3.773 GeV, the $D^0\bar{D}^0$ pairs can be coherently produced through $\ee \to (D^0\bar{D}^0)_{\text{C}=-}$, and above the threshold of $D^{*0}\bar{D}^{*0}$, $e.g.$, at 4.030 GeV, the reactions of $\ee \to D^0\bar{D}^{*0} \to \pi^0 (D^0\bar{D}^0)_{\text{C}=-}$ or $\gamma (D^0\bar{D}^0)_{\text{C}=+}$, as well as $\ee \to D^{*0}\bar{D}^{*0} \to \pi^0\pi^0 (D^0\bar{D}^0)_{\text{C}=-}$, $\gamma\gamma (D^0\bar{D}^0)_{\text{C}=-}$ or $\gamma \pi^0 (D^0\bar{D}^0)_{\text{C}=+}$ can be utilized.
Therefore, useful constraints on $D^0$-$\bar{D}^0$ mixing and CP-violating parameters may be obtained in the respective decays of the correlated $D^0$ and
$\bar{D}^0$ events~\cite{Xing97}.
For example, the mixing rate $R_M = (x^2 + y^2)/2$ via the same charged final states $(K^\pm\pi^\mp)(K^\pm\pi^\mp)$ or $(K^\pm\ell^\mp\nu)(K^\pm\ell^\mp\nu)$ with a sensitivity of $10^{-5}$ using 1 ab$^{-1}$ of data at 3.773 GeV. With approximately 5 ab$^{-1}$ of data at 4.030 GeV, the sensitivity for the mixing parameters $x$ and $y$ is expected to be below 0.02\%, whereas the measurements of $|q/p|$ and $\arg(q/p)$ have the potential to achieve precisions of 1.5\% and $1.2^\circ$, respectively~\cite{Bondar:2010qs}.
These measurements are complementary to the precision measurements at Belle II and the upgraded LHCb. Furthermore, the decay mode $( D^0\bar{D}^0 )_{\text{CP}=\pm} \to ( f_1 f_2 )_{\text{CP}=\mp}$ is CP-forbidden unless a CP violation exists. The rate of a pair of CP-even final states $f_+$ (e.g., $\pi^+\pi^-$) can be described with $\Gamma^{++}_{D^0\bar{D}^0} = \left[ \left(x^2 + y^2\right)\left(\cosh^2 a_m - \cos^2 \phi\right) \right] \Gamma^2(D \to f_+)$, where $\phi = \arg(p/q)$, $R_\text{M} = |p/q|$, and $a_m = \log R_{\text{M}}$~\cite{Atwood:2002ak}.

CPT conservation is a cornerstone of the SM and many of its extensions. However, CPT violations can theoretically arise in some models such as string theory or extra-dimensional models with Lorentz-symmetry violations. Therefore, observing T violations without assuming CPT conservation is critical ~\cite{Shi:2016bvo}. Studies on the time evolution of CP-correlated $D^0$-$\bar{D}^0$ states at the STCF complement CPT-violation studies at super-$B$ factories and LHCb experiments~\cite{Kostelecky:2001ff}. However, achieving this requires asymmetric $\ee$ collisions owing to the low momentum of the produced $D$ mesons in symmetric collisions.

The quantum correlations of $D^0\bar{D}^0$ pairs also enable the probing of the amplitudes of $D^0$ decays and the determination of the strong phase difference between the Cabibbo-favored and doubly Cabibbo-suppressed amplitudes~\cite{Wilkinson:2021tby}. This is important for understanding the nonperturbative QCD effects, extracting the CKM angle $\gamma$, and relating the measured mixing parameters in hadronic decays ($x'$, $y'$) to the mass and width-difference parameters ($x$, $y$)~\cite{Amhis:2019ckw}.

The measurements of the CKM unitary triangle (UT) angles $\alpha$, $\beta$, and $\gamma$ in $B$ decays are crucial tests of CKM unitarity and are essential for probing potential CP violations beyond the SM. Any discrepancy between measurements involving tree- and loop-dominated processes in the UT signals the presence of new heavy degrees of freedom contributing to these loops. Among the three CKM angles, $\gamma$ is particularly significant because it is the only CP-violating observable that can be determined through tree-level decays. The precise measurement of $\gamma$ will be a key focus for both the LHCb upgrade(s) and Belle II experiments.

The most precise method for determining $\gamma$ depends on the interference between the $B^{+}\to\bar{D}^{0}K^{+}$ and $B^{+}\to D^{0}K^{+}$ decay modes~\cite{GLW, ADS, GGSZ}. In future, the statistical uncertainties in these measurements will be significantly reduced by utilizing the vast $B$ meson samples collected by the LHCb and Belle II. However, limited knowledge of the strong phases in $D$ decays systematically limits the overall sensitivity. For example, a dataset of 20 fb$^{-1}$ at 3.773 GeV from BESIII is expected to result in a systematic uncertainty of approximately 0.4$^\circ$ for $\gamma$ measurements~\cite{Ablikim:2019hff}. To match the anticipated statistical uncertainty of less than 0.4$^\circ$ in LHCb upgrade II, STCF will play a critical role by providing constraints that reduce the systematic uncertainty from $D$ strong-phase measurements to less than 0.1$^\circ$, thereby enabling detailed comparisons of $\gamma$ results across different decay modes.

\subsection{Rare and forbidden decays}

STCF, endowed with its high luminosity, clean collision environment, and superior detector capabilities, holds considerable promise for searching for rare and forbidden $D$-meson decays and may serve as a valuable probe for new physics beyond the SM. These searches can be categorized into three main groups.
\begin{itemize}
\item{Flavor-Changing Neutral Current (FCNC) Decays}:
These include decays such as $D^{0(+)} \to \gamma V^{0(+)}$, $D^0 \to \gamma\gamma$, $D^0 \to \ell^+\ell^-$, and $D \to \ell^+\ell^- X$, as well as $D \to \nu \overline{\nu} X$. These decays are allowed in the SM; however, they are highly suppressed and proceed through loop diagrams. New physics can enhance the decay rates, making them significant targets for investigation.
\item{Lepton Flavor Violating (LFV) Decays}:
Such decays involve processes such as $D^0 \to \ell^+ \ell^{\prime -}$ and $D \to \ell^+\ell^{\prime -} X$ (for $\ell \neq \ell^\prime$), which are forbidden in the SM. The observation of neutrino oscillations has confirmed LFV decays in the lepton sector, making the search for LFV phenomena in the charm-quark sector a meaningful endeavor.
\item{Lepton Number Violating (LNV) Decays}:
These decays, such as $D^+ \to \ell^+\ell^{\prime +} X^-$ and $D^+_s \to \ell^+\ell^{\prime +} X^-$ (for either $\ell = \ell^\prime$ or $\ell \neq \ell^\prime$), are also forbidden in the SM. If neutrinos are Majorana particles, LNV decays may occur, further motivating the search for these processes.
\end{itemize}

Although the FCNC decays of $D$ mesons are allowed and dominated by long-distance dynamics in the SM, their decay rates are still very small. For example, the predicted BFs for $D^0 \to \gamma\gamma$ and $D^0 \to \mu^+\mu^-$ are approximately $1 \times 10^{-8}$ and $3 \times 10^{-13}$, respectively~\cite{Burdman}; however, new physics can significantly enhance these values~\cite{Golowich}. The current experimental limits for these BFs are $8.5 \times 10^{-7}$ and $6.2 \times 10^{-9}$ for $D^0 \to \gamma\gamma$ and $D^0 \to \mu^+\mu^-$, respectively~\cite{PDG}. 
Semileptonic decays, such as $D^0 \to \pi^+ \pi^- \mu^+ \mu^-$, $D^0 \to K^+ K^- \mu^+ \mu^-$, and $D^0 \to K^- \pi^+ \mu^+ \mu^-$ have been observed in LHCb experiments with BFs at the $10^{-7}$ level \cite{PDG}. These observations suggest nontrivial contributions from the complicated long-distance effects.

At the STCF, the study of di-electron modes $D \to e^+e^- X$ is particularly advantageous, providing sensitivities ranging from $10^{-8}$ to $10^{-9}$ for the invariant mass $m_{e^+e^-}$ in ranges free from long-distance contributions~\cite{TheBESIIICollaboration2018a}. The clean background environment of STCF makes it competitive for channels containing neutral final states, such as photons and $\pi^0$. Moreover, the STCF offers the best constraints on the upper limits of the BFs for rare $D$ decays involving neutrinos, such as $D^0 \to \pi^0 \nu \overline{\nu}$~\cite{BESIII:2021slf} and $D^0 \to \gamma \nu \overline{\nu}$.

To date, no evidence has been observed for the forbidden $D_{(s)}$-meson decay involving the LFV decays, LNV decays, or both. The current limits on the LFV decays are generally set at $10^{-6}$--$10^{-5}$, except for $D^0 \to \mu^\pm e^\mp$, where the limit is $1.3 \times 10^{-8}$~\cite{PDG}. The STCF is poised to provide even more stringent limits on these LFV and LNV decay modes, with sensitivities ranging from $10^{-8}$ to $10^{-9}$ or better, leveraging its clean environment and precise charge discrimination.

\subsection{Charmed meson spectroscopy}

The STCF will serve as an exceptional platform for the exploration of charmed meson production and spectroscopy. Presently, all $1S$ and $1P$ $D_{(s)}$ states have been confirmed experimentally~\cite{Chen:2016spr}. However, the experimental landscape remains incomplete, with many excited states predicted from QCD-derived effective models yet to be observed. In addition, numerous excited open-charm states have been reported, and their nature remains contentious, with some candidates potentially representing exotic mesons.
For example, the narrow $D^*_{sJ}(2632)$ state observed by the SELEX collaboration has not been corroborated by subsequent searches conducted at CLEO, BaBar, and FOCUS, leading to conflicting interpretations. Similarly, the unexpectedly low masses of $D_{s0}^{*}(2317)$ and $D_{s1}(2460)$ have sparked various explanations, including the possibility that these are $D^{(*)}K$ molecular states~\cite{Guo:2017jvc}. Strong $S$-wave$ D^{(*)}K$ scattering is suggested to contribute to the observed mass reduction. Therefore, more comprehensive studies on open-charm meson spectroscopy are urgently required.

At the STCF, excited charmed meson states $D^{**}$ can be copiously produced through direct electron-positron annihilation processes such as $\ee \to D^{**}\bar{D}^{(*)}(\pi)$ across an energy range of 4.1--7.0 GeV. These states can then be studied via hadronic or radiative decays into lower-lying open-charm states~\cite{Kato:2018ijx}. Such systematic investigations at the STCF will provide essential data to explore nonperturbative QCD dynamics in the charm sector and test various theoretical models.

\section{Charmed baryon}

Theoretical studies on the weak decay of charmed baryons increased in the early 1990s, which subsequently decreased. However, experimental studies on charmed baryon decay have increased, particularly regarding the hadronic weak decay of $\Lambda_c^+$ at BESIII~\cite{Li:2021iwf}.
Another notable breakthrough was the reordering of the lifetime hierarchy of charmed baryons, as determined by the LHCb collaboration. The new hierarchy established by LHCb is $\tau_{\Xi_c^+} > \tau_{\Omega_c^0} > \tau_{\Lambda_c^+} > \tau_{\Xi_c^0}$ from the previous order $\tau_{\Xi_c^+}>\tau_{\Lambda_c^+}>\tau_{\Xi_c^0}>\tau_{\Omega_c^0}$~\cite{LHCb:2018nfa,LHCb:2019ldj,LHCb:2021vll,Cheng:2021vca}. These advancements sparked renewed theoretical interest in the study of hadronic weak decays of singly charmed baryons, as they challenged previous expectations and necessitated a deeper understanding of the underlying dynamics.

Charmed baryon spectroscopy is particularly valuable for studying the behavior of light quarks in the presence of heavy quarks. Over the past decade, numerous new excited states of charmed baryons have been discovered in experiments such as BaBar, Belle, CLEO, and LHCb. Both $B$ decay and $e^+e^-$ annihilation into $c\bar{c}$ pairs are rich sources of charmed baryons, and considerable effort has been devoted to identifying the quantum numbers of these new states and understanding their properties.

\subsection{Hadronic weak decays}

The BFs for Cabibbo-favored two-body decays of $\Lambda_c^+$ are summarized in Table~\ref{tab:BRs}. Among them, a few channels are still elusive from experimental measurements because amplitude analyses must be conducted in the same manner as those in the studies of $\Lambda_c^+\to \Lambda \rho^+$~\cite{BESIII:2022udq} and $\Sigma^{*+}\eta$~\cite{BESIII:2024mbf}.
Several decay modes such as $\Sigma^+\phi$, $\Xi^{(*)}K^{(*)+}$ and $\Delta^{++}K^-$ proceed exclusively through $W$-exchange processes. The experimental observations of these decays highlight the significant role of $W$-exchange, which is not subject to color suppression. Belle~\cite{Zupanc} and BESIII~\cite{BES:pKpi} measured the absolute BF for the decay $\Lambda_c^+\to pK^-\pi^+$. The PDG reported a new average of $(6.24 \pm 0.28)\%$ for the benchmark mode \cite{PDG}. In addition, the doubly Cabibbo-suppressed decay $\Lambda_c^+\to pK^+\pi^-$ was observed by Belle \cite{Belle:DCS} and LHCb \cite{LHCb:DCS}, emphasizing the need to search for two-body doubly Cabibbo-suppressed modes, such as $pK^{0(*)}$ and $nK^{+(*)}$.

\begin{table*}[tp]
\caption{The measured BFs for Cabibbo-favored two-body decays of $\Lambda_c^+$ (in percentage) from PDG~\cite{PDG}.} \label{tab:BRs}
\begin{center}
\begin{tabular}{lc| lc|lc}
\hline\hline Mode & $\B$ & Mode & $\B$ & Mode & $\B$  \\
\hline
$\Lambda \pi^+$ & 1.29$\pm$0.05 & $ \Lambda \rho^+$ & $4.0\pm0.5$ & $ \Delta^{++}K^-$  &  $1.76\pm0.09$\\
\hline
$ \Sigma^0 \pi^+$ & 1.27$\pm$0.06 & $ \Sigma^0 \rho^+$  & & $ \Sigma^{*0} \pi^+$ & \\
\hline
$ \Sigma^+ \pi^0$ & 1.24$\pm$0.09 & $ \Sigma^+ \rho^0$  & $<1.7$ & $ \Sigma^{*+} \pi^0$& \\
\hline
$ \Sigma^+ \eta$ & 0.32$\pm$0.05 & $ \Sigma^+ \omega$  &  1.69$\pm$0.20 & $ \Sigma^{*+}\eta$ & $0.91\pm0.20$ \\
\hline
$ \Sigma^+ \eta^\prime$ & 0.41$\pm$0.08 & $ \Sigma^+ \phi$ & 0.39$\pm$0.05 & $ \Sigma^{*+} \eta^\prime$ &\\
\hline
$ \Xi^0 K^+$ & 0.55$\pm$0.07 & $ \Xi^0 K^{*+}$ &  & $ \Xi^{*0}K^+$  & 0.43$\pm$0.09 \\
\hline
$ p K_S$ &  1.59$\pm$0.07 & $ p \bar K^{*0}$  & 1.39$\pm$0.07 & $ \Delta^+\bar K^0$ &  \\ 
\hline \hline
\end{tabular}
\end{center}
\end{table*}

Various theoretical approaches to the weak decay of heavy baryons have been explored, including the current algebra method, factorization approach, pole model, relativistic quark models, quark diagram scheme, and SU(3) flavor symmetry. Generally, apart from the current algebra, most of these models predict decay rates that are lower than the experimental results. Furthermore, decay asymmetries in two-body hadronic weak decays of charmed baryons are of interest. These asymmetries are defined as
$\alpha\equiv\frac{2{\rm Re}(s^* p)}{|s|^2+|p|^2}$, where
$s$ and $p$ represent the parity-violating $s$-wave and parity-conserving $p$-wave amplitudes during decay, respectively. The covariant quark model and its variants, as well as the pole model, predicted a positive decay asymmetry $\alpha$ for both $\Lambda_c^+\to \Sigma^+\pi^0$ and $\Lambda_c^+\to \Sigma^0\pi^+$, whereas CLEO \cite{CLEO:alpha} measured the value for $\Sigma^+\pi^0$ as $-0.45 \pm 0.31 \pm 0.06$. In contrast, the current algebraic approach consistently predicted a negative decay asymmetry for these modes, with values ranging from $-0.49$ to $-0.76$~\cite{CT93,Verma98,Zen:1993,Datta}. The sign of $\alpha_{\Sigma^+\pi^0}$ was finally confirmed by BESIII, which measured $\alpha_{\Sigma^+\pi^0}=-0.57 \pm 0.12$ \cite{BES:deasy}, thereby verifying the earlier CLEO result.

The absolute BFs for $\Xi_c^0 \to \Xi^-\pi^+$ and $\Xi_c^+ \to \Xi^-\pi^+\pi^+$ were measured by Belle \cite{Belle:Xic0,Belle:Xic+} as $\B(\Xi_c^0\to \Xi^-\pi^+) = (1.80 \pm 0.50 \pm 0.14)\%$  and $\B(\Xi_c^+\to \Xi^-\pi^+\pi^+) = (2.86 \pm 1.21 \pm 0.38)\%$.
These measurements enable inferences regarding the BFs for the other $\Xi_c^0$ and $\Xi_c^+$ decays. However, absolute BFs were not measured for $\Omega_c^0$. The hadronic weak decays of $\Omega_c^0$ were recently studied in detail in Ref.~\cite{Dhir}, where most decay channels proceeded predominantly through $W$-exchange diagrams.

With the improved precision expected at STCF, the nonleptonic decay modes of $\Lambda_c^+$ and $\Xi_c^{+,0}$ may be measured more accurately. Of particular interest are the decay asymmetries $\alpha$ in various charm baryon decays and the absolute BFs of $\Omega_c^0$ decays.

Heavy-flavor-conserving hadronic decays of charmed baryons form a special class of weak decays that can be studied more reliably. These decays involve weak interactions of only the light quarks within the charmed baryon, whereas the heavy quark acts as a spectator. Examples of such decays include singly Cabibbo-suppressed decays, $\Xi_c \to \Lambda_c^+ \pi$, and $\Omega_c \to \Xi'_c \pi$. The synthesis of heavy quarks and chiral symmetries provides a natural framework for analyzing these reactions~\cite{ChengHFC}.
Theoretical predictions of the BFs for $\Xi_c^0 \to \Lambda_c^+ \pi^-$ and $\Xi_c^+ \to \Lambda_c^+ \pi^0$ are typically in the range of $10^{-3}$--$10^{-4}$~\cite{ChengHFC}. The LHCb experiment reported the first measurement of $\Xi_c^0 \to \Lambda_c^+ \pi^-$ with a BF of $(0.55 \pm 0.02 \pm 0.18)\%$~\cite{Aaij:2020wtg}, which was higher than the theoretical predictions. Afterwards, the LHCb measurement was confirmed by Belle with the BF
result of $(0.54\pm 0.05\pm 0.05\pm 0.12)\%$~\cite{Belle:2022kqi}.
The STCF is poised to make significant contributions to this field by verifying the LHCb result for $\Xi_c^0 \to \Lambda_c^+ \pi^-$ and searching for a charm-flavor-conserving weak decay $\Xi_c^+ \to \Lambda_c^+ \pi^0$. 

\subsection{Semileptonic decays}

\begin{table*}[tp]
\caption{Determined BFs for semileptonic decays of $\Lambda_c^+$ (in units of 0.1\%) with relative precisions in parenthesis. In addition, the projected statistical precisions at the STCF are provided.} \label{tab:LcSL}
\begin{center}
\begin{tabular}{lcc|lcc}
\hline\hline Mode & BESIII & STCF & Mode & BESIII & STCF  \\
\hline
\hline
$\Lambda e^+ \nu_e$ &  35.6$\pm$1.3 (3.6\%)~\cite{BESIII:2022ysa} &  0.2\%  & $\Lambda \mu^+ \nu_\mu$  & 34.8$\pm$1.7  (4.9\%)~\cite{BESIII:2023jxv}  & 0.3\%  \\
 $p K^- e^+ \nu_e$ & $0.88\pm0.18$ (20\%)~\cite{BESIII:2022qaf} & 1.3\% & $n e^+ \nu_e$ & $3.57\pm0.37$(10\%)~\cite{BESIII:2024mgg} & 0.6\% \\ \hline \hline
\end{tabular}
\end{center}
\end{table*}

Exclusive semileptonic decays of charmed baryons, such as $\Lambda_c^+ \to \Lambda e^+(\mu^+) \nu_{e(\mu)}$,  $\Lambda_c^+ \to p K^- e^+ \nu_{e}$, $\Lambda_c^+ \to n e^+ \nu_{e}$, $\Xi_c^+ \to \Xi^0 e^+ \nu_e$, and $\Xi_c^0 \to \Xi^- e^+ \nu_e$, have been observed experimentally. The rates of these decays depend critically on the form factors that describe the transition-matrix elements between the initial and final baryon states.  Table~\ref{tab:LcSL} lists the determined BFs for the semileptonic decays of $\Lambda_c^+$ as well as the projected statistical precision at STCF. Various theoretical models have been adopted to evaluate these form factors, including the nonrelativistic quark model~\cite{Marcial,Singleton,CT96,Pervin}, MIT bag model~\cite{Marcial}, relativistic quark models~\cite{Ivanov96,Gutsche,Faustov:semi}, light-front quark model~\cite{Luo}, QCD sum rules~\cite{Carvalho,Huang,Azizi}, and LQCD calculations~\cite{Meinel:LamcLam,Meinel:Lamcn}.

Early predictions of $\mathcal{B}(\Lambda_c^+ \to \Lambda e^+ \nu_e)$ were typically smaller than the experimental value of $(3.6 \pm 0.4)\%$ measured by BESIII~\cite{BESIII:Lambdaenu}. However, recent LQCD calculations~\cite{Meinel:LamcLam} have shown good agreement with experimental data for $\Lambda_c^+ \to \Lambda e^+ \nu_e$ and $\Lambda_c^+ \to \Lambda \mu^+ \nu_\mu$.

The STCF will provide a unique opportunity to study the semileptonic decays of various charmed baryons, including $\Lambda_c^+ \to \Lambda^* \ell^+ \nu_\ell$ and $\Lambda_c^+ \to N^* \ell^+ \nu_\ell$, as well as the $\Xi_c^{+,0}$ and $\Omega_c^0$ semileptonic decays, which will enable detailed discrimination among different form-factor models.

\subsection{Electromagnetic and weak radiative decays}

Electromagnetic transitions between two charmed baryons play a crucial role in understanding the dynamics of these particles and their underlying symmetrical structures. These decays include:
(i) $\Sigma_c^+ \rightarrow \Lambda_c^+  \gamma$, $\Xi^\prime_c \rightarrow \Xi_c  \gamma$,
(ii) $\Sigma^{\ast+}_c \rightarrow \Lambda_c + \gamma$, $\Xi^\ast_c \rightarrow \Xi_c + \gamma$,
(iii) $\Sigma^\ast_c \rightarrow \Sigma_c  \gamma$, $\Xi^\ast_c \rightarrow \Xi^\prime_c  \gamma$, $\Omega^{\ast 0}_c \rightarrow \Omega^{0}_c  \gamma$, and
(iv) $\Lambda^+_c(2595, 2625) \to \Lambda_c \gamma$, $\Xi_c(2790, 2815) \to \Xi_c \gamma$.
The decay modes $\Xi'^0_c \to \Xi_c^0 \gamma$, $\Xi'^+_c \to \Xi^+_c \gamma$, and $\Omega_c^{*0} \to \Omega_c^0 \gamma$ have been observed experimentally.
Table \ref{tab:em} summarizes the theoretical predictions of the electromagnetic decay rates of $s$-wave charmed baryons using heavy hadron chiral perturbation theory (HHChPT) at different orders: leading order (LO) \cite{Cheng97,Cheng93}, next-to-leading order (NLO)~\cite{Jiang}, and next-to-next-to-leading order (NNLO)~\cite{Wang:2018}.
The significant differences in the predicted decay rates at the NLO from those at the LO and NNLO for the modes $\Sigma_c^{*+} \to \Lambda_c^+ \gamma$, $\Sigma_c^{*++} \to \Sigma_c^{++} \gamma$, and $\Xi^{*+}_c \to \Xi_c^+ \gamma$ are unclear.
All the HHChPT approaches should agree with each other regarding the lowest order of chiral expansion if the coefficients are derived from the nonrelativistic quark model. 
Clarifying these discrepancies requires precise experimental measurements in which the STCF is well positioned for delivery.

\begin{table*}[tp]
\centering
\normalsize
\caption{Electromagnetic partial width (in units of keV) of $s$-wave charmed
baryons in heavy hadron chiral perturbation theory in (i) LO \cite{Cheng97,Cheng93}, (ii) NLO \cite{Jiang}, and (iii) NNLO  \cite{Wang:2018}.} \label{tab:em}
\begin{center}
\begin{tabular}{cccccc}
\hline\hline & $\Sigma^+_c\to \Lambda_c^+\gamma$ & $\Sigma_c^{*+}\to\Lambda_c^{+}\gamma$ & $\Sigma_c^{*++}\to\Sigma_c^{++}\gamma$ & $\Sigma_c^{*0}\to\Sigma_c^0\gamma$ &  $\Xi'^+_c\to\Xi_c^+\gamma$ \\ \hline
 (i) & 91.5 & 150.3 & 1.3 & 1.2 & 19.7  \\
 (ii) & 164.2 & 893.0 & 11.6 & 2.9 & 54.3 \\
 (iii) & 65.6 & 161.8 & 1.2 & 0.49 & 5.4 \\
 \hline\hline & $\Xi^{*+}_c\to\Xi_c^+\gamma$ & $\Xi^{*0}_c\to\Xi_c^0\gamma$ & $\Xi'^0_c\to\Xi_c^0\gamma$ & $\Omega_c^{*0}\to\Omega_c^0\gamma$ & \\ \hline
 (i) & 63.5 & 0.4 & 1.0 & 0.9 & \\
 (ii) &  502.1 & 0.02 & 3.8 & 4.8 &\\
 (iii) &  21.6 & 0.46 & 0.42 & 0.32 & \\
 \hline \hline
\end{tabular}
\end{center}
\end{table*}

Belle observed the electromagnetic decay of the orbitally excited charmed baryons $\Xi_c(2790)$ and $\Xi_c(2815)$~\cite{Belle:charme.m.}. The partial widths of $\Xi_c(2815)^0 \to \Xi_c^0 \gamma$ and $\Xi_c(2790)^0 \to \Xi_c^0 \gamma$ were $320 \pm 45^{+45}_{-80}$ keV and $\sim 800$ keV, respectively. However, no signals were observed for the analogous decay of the charged states $\Xi_c(2815)^+$ and $\Xi_c(2790)^+$.

In addition, weak radiative decays, such as $\Lambda_c^+ \to \Sigma^+ \gamma$ and $\Lambda_c^+ \to p \gamma$, can occur through bremsstrahlung processes. The former proceeds via the Cabibbo-favored transition $cd \to us\gamma$, whereas the latter involves the Cabibbo-suppressed transition $cd \to ud\gamma$. The BF for $\Lambda_c^+ \to \Sigma^+ \gamma$ is estimated to be on the order of $10^{-4}$ \cite{Chengweakrad}.
BESIII reported an upper limit of $\B(\Lambda_c^+ \to \Sigma^+ \gamma)<4.4\times 10^{-4}$~\cite{BESIII:2022aok}, which is on the order of the theoretical estimation.
More studies on these decays at the STCF will be instrumental in identifying the BFs
and testing theoretical predictions, thereby advancing our understanding of strong interaction dynamics within charmed baryons.

\subsection{CP violation}

The CKM matrix contains a complex phase that allows for CP-violating phenomena in the SM. Although CP violations have been extensively studied in meson systems, such effects may be observed in the decay of charmed baryons, although the predicted CP-violating asymmetries tend to be small.
Recently, the search for CP violations in charmed baryon decays has increased, driven by the availability of large samples of $\Lambda_c^+$ baryons collected in the BESIII, LHCb, and Belle (II) experiments. For the two-body decays of $\Lambda_c^+$, a CP violation can be probed by measuring the CP-violating asymmetry ${\cal A}=(\alpha+\bar\alpha)/(\alpha-\bar\alpha)$, where the decay asymmetries $\alpha$ and $\bar\alpha$ correspond to the decays $\Lambda_c^+$ and $\bar{\Lambda}_c^-$, respectively. For example, the FOCUS experiment measured ${\cal A}$ in $\Lambda_c^+\to \Lambda\pi^+$ and $\bar\Lambda_c^-\to\bar\Lambda \pi^-$ decays to be $-0.07\pm0.19\pm0.24$ \cite{Link:2005ft}.
The STCF will offer a significant enhancement in the sensitivity of searches for CP violation by combining "single tag" $\Lambda_c^+$ data~\cite{Ablikim:2019zwe} with "double tag" $\Lambda_c^+\bar{\Lambda}_c^-$ data, where the pairs of $\Lambda_c^+\bar{\Lambda}_c^-$ are quantum-correlated with respect to their spins aligned to the initial transverse polarization of the virtual photon. Using polarized beams, the STCF can achieve enhanced sensitivities to decay asymmetries and CP violations, given the known direction of the spin orientation of the produced $\Lambda_c^+$~\cite{Bondar:2019zgm}.

For three-body decays, the LHCb measured the difference in the CP asymmetries of $\Lambda_c^+ \to p K^+ K^-$ and $\Lambda_c^+ \to p \pi^+ \pi^-$, $\Delta A_{CP}$, as $(0.30 \pm 0.91 \pm 0.61)\%$. 
This result, which is consistent with zero, is compared with a generic SM prediction of a fraction of 0.1\% \cite{Aaij:2017xva,Bigi:2012ev}. 
An increase in statistics by at least a factor of 100 is required to probe the size of the CP asymmetry in the SM.
The four-body hadron final states of $\Lambda_c^+$ decay, such as $\Lambda_c^+\to pK^-\pi^+\pi^0$, $\Lambda_c^+\to\Lambda\pi^+\pi^+\pi^-$, and $\Lambda_c^+\to pK_S\pi^+\pi^-$, offer opportunities to explore CP violations through T-odd observables. The STCF, characterized by its high luminosity, broad center-of-mass energy coverage, abundant production of charmed baryons, and clean experimental environment, is an ideal facility for such studies.
Fast simulations indicate that with 1 ab$^{-1}$ of $e^+e^-$ annihilation data at $\sqrt{s}=4.64$ GeV, as expected at the STCF, a sensitivity of (0.25--0.50)\% can be achieved for the aforementioned three decay modes. This sensitivity is sufficient to measure nonzero CP-violating asymmetries as large as 1\%~\cite{Shi:2019vus}.

\subsection{Spectroscopy}

Table~\ref{tab:3and6} lists the observed antitriplet ($\mathbf{\bar{3}}$) and sextet ($\mathbf{6}$) states of the charmed baryons, along with their mass differences. These states include $\Lambda_c^+$, $\Sigma_c$, $\Xi^{(\prime)}_c$, and $\Omega_c^0$ baryons with various spin-parity ($J^P$) configurations.
The highest state in the $\Lambda_c$ family, $\Lambda_c(2940)^+$, was initially discovered decaying into $D^0p$  by BaBar~\cite{BaBar:Lamc2940}, and has been a subject of debate regarding its $J^P$ assignment~\cite{Cheng:2015}. Recent studies have suggested that $J^P = \frac{3}{2}^-$ based on LHCb data~\cite{LHCb:Lambdac2880}; however, arguments also suggest that $J^P = \frac{1}{2}^-(2P)$ based on Regge analysis~\cite{Cheng:Omegac}.
However, Ref.~\cite{Luo:2019qkm} proposed that $\Lambda_c(2940)^+$ is a $\frac32^-(2P)$ state and another state $\frac12^-(2P)$ is heavier than $\Lambda_c(2P, 3/2^-)$. 
In addition, LHCb observed five narrowly excited $\Omega_c$ states that decayed into $\Xi_c^+ K^-$: $\Omega_c(3000)$, $\Omega_c(3050)$, $\Omega_c(3066)$, $\Omega_c(3090)$, and $\Omega_c(3119)$~ \cite{LHCb:Omegac}. The identification of their spin-parity quantum numbers remains an interesting research subject.

STCF is well suited for studying the spectroscopy of singly charmed baryons in the energy range of $4.6-7.0$ GeV, enabling the exploration of their possible structures and spin-parity quantum-number assignments. Thus, investigating the newly observed $\Omega_c$ excited resonances. If the energy region is extended to more than 7.4 GeV, the production of doubly charmed baryons such as $\Xi^{++}_{cc}$ becomes feasible, opening up new avenues for studying these new states in more detail. 

\begin{table*}[tp]
\begin{center}
\normalsize
\caption{Charmed baryon spectroscopy of antitriplet and sextet states.
Mass differences are given as $\Delta m_{\Xi_c\Lambda_c}\equiv m_{\Xi_c}-m_{\Lambda_c}$, $\Delta m_{\Xi'_c\Sigma_c}\equiv m_{\Xi'_c}-m_{\Sigma_c}$, $\Delta m_{\Omega_c\Xi'_c}\equiv m_{\Omega_c}-m_{\Xi'_c}$ in units of MeV. } \label{tab:3and6}

\begin{tabular}{c| ccc } \hline\hline
  & $J^P(nL)$ & States & Mass difference  \\
 \hline
 ${\bf \bar 3}$ & ${1\over 2}^+(1S)$ &  $\Lambda_c(2287)^+$, $\Xi_c(2470)^+,\Xi_c(2470)^0$ & $\Delta m_{\Xi_c\Lambda_c}=183$   \\
 & ${1\over 2}^-(1P)$ &  $\Lambda_c(2595)^+$, $\Xi_c(2790)^+,\Xi_c(2790)^0$ & $\Delta m_{\Xi_c\Lambda_c}=198$  \\
 & ${3\over 2}^-(1P)$ &  $\Lambda_c(2625)^+$, $\Xi_c(2815)^+,\Xi_c(2815)^0$ & $\Delta m_{\Xi_c\Lambda_c}=190$  \\
 & ${3\over 2}^+(1D)$ &  $\Lambda_c(2860)^+$, $\Xi_c(3055)^+,\Xi_c(3055)^0$ & $\Delta m_{\Xi_c\Lambda_c}=201$  \\
 & ${5\over 2}^+(1D)$ &  $\Lambda_c(2880)^+$, $\Xi_c(3080)^+,\Xi_c(3080)^0$ & $\Delta m_{\Xi_c\Lambda_c}=196$  \\
 \hline
 ${\bf 6}$ & ${1\over 2}^+(1S)$ &  $\Omega_c(2695)^0$, $\Xi'_c(2575)^{+,0},\Sigma_c(2455)^{++,+,0}$ & $\Delta  m_{\Omega_c\Xi'_c}=119$, $\Delta m_{\Xi'_c\Sigma_c}=124$  \\
 & ~${3\over 2}^+(1S)$~ &  $\Omega_c(2770)^0$, $\Xi'_c(2645)^{+,0},\Sigma_c(2520)^{++,+,0}$ &  $\Delta m_{\Omega_c\Xi'_c}=120$, $\Delta m_{\Xi'_c\Sigma_c}=128$ \\
 \hline\hline
\end{tabular}
\end{center}
\end{table*}

\section{Summary}

In conclusion, the STCF will be an ideal facility for investigating charm physics through the unique pair production of charmed hadrons near the threshold. Specifically, it will enable the precise determination of charm mixing and CP-violating parameters, as well as the strong phase of neutral $D$ mesons, by utilizing the quantum correlations of the $D^0\bar{D}^0$ pair system. 
With approximately 5 ab$^{-1}$ of data at 4.030 GeV, the expected sensitivity of the mixing parameters ($x$, $y$) is projected to reach an impressive 0.02\%, whereas those for $|q/p|$ and $\arg(q/p)$ are expected to be 1.5\% and $1.2^\circ$, respectively. 
Additionally, the $D$ strong-phase measurements at the STCF will be critical for reducing the uncertainty of the CP-violating phase $\gamma$ to below 0.1$^\circ$, enabling detailed comparisons of $\gamma$ results from different analysis channels in the future LHCb upgrade II. These measurements will complement studies at the LHCb and Belle II, providing vital insights into the mechanism of CP violations. Furthermore, the leptonic decay of charmed mesons and baryons will be explored with state-of-the-art precision, offering stringent constraints on the CKM matrix, clean probes for strong-force dynamics, rigorous LFU tests, and potential pathways to new physics discoveries.

In the energy region above the charmed baryon threshold, the STCF will facilitate extensive studies on the production and decay properties of (excited) charmed baryons, which still lack comprehensive experimental data. The absolute measurements of the semileptonic and nonleptonic decays of the $\Lambda_c^+$, $\Xi_c^{+,0}$, and $\Omega_c^0$ baryons will be significantly improved. A key focus will be placed on decay asymmetries (such as $\alpha$) in various charmed baryon decays, which will also serve as a potential avenue for searching for CP violation. The absolute BFs of $\Omega_c^0$ decays will also be measured with unprecedented precision. Moreover, the search for rare and forbidden decays of charmed hadrons can yield sensitivity improvements of up to two orders of magnitude, further advancing the search for new physics.

Many of these crucial measurements will benefit from the clean reaction environment and well-constrained kinematics at the STCF. Additionally, the detector designed for optimal performance, particularly in charged-particle identification, low-momentum charged-particle detection, and photon measurement, will be advantageous. The unprecedented precision achievable with high-statistics charm data will deepen our understanding of the challenges of the SM and may provide new evidence or solutions. Thus, the STCF will play a pivotal role in advancing the high-intensity frontier of elementary particle physics on a global scale.

\end{document}